\newcommand{\version}{February 6, 2007}

\documentclass[12pt]{article}
\usepackage{bbm}
\usepackage{a4,amsthm,amsfonts,latexsym,amssymb}

\swapnumbers
\theoremstyle{plain}
\newtheorem{thm}{THEOREM}
\newtheorem{cl}[thm]{COROLLARY}
\newtheorem{lem}[thm]{LEMMA}
\newtheorem{define}[thm]{DEFINITION}
\newtheorem{proposition}[thm]{PROPOSITION}

\theoremstyle{definition}

\newcommand{\beq}{\begin{equation}}
\newcommand{\eeq}{\end{equation}}
\def\beqa{\begin{eqnarray}}
\def\eeqa{\end{eqnarray}}

\newcommand{\C}{{\mathbb C}}
\newcommand{\W}{{\mathcal W}}
\newcommand{\sdird}{\mbox{$\frac{1}{\sqrt{d}}$}}
\newcommand{\R}{{\mathbb R}}
\newcommand{\Z}{{\mathbb Z}}
\newcommand{\one}{{\mathbbm 1}}
\newcommand{\Tr}{{\rm Tr}}
\newcommand{\Ww}{{\mathbb W}}
\newcommand{\T}{{\mathbb T}}

\newcommand{\G}{{\mathcal G}}

\newcommand{\Hh}{{\mathcal H}}

\newcommand{\eps}{\varepsilon}

\date{\small\version}

\begin{document}
\markboth{\scriptsize{Qdits BHN  \version}}{\scriptsize{Qdits BHN
\version}}

\title{
\vspace{-80pt}
\begin{flushright}
{\small UWThPh-2006-23}
\vspace{30pt}
\end{flushright}
\bf{A special simplex in the state space for entangled qudits}}
\author{\vspace{8pt} Bernhard Baumgartner$^1$ , Beatrix Hiesmayr$^2$ , Heide Narnhofer$^3 $\\
\vspace{-4pt}\small{Institut f\"ur Theoretische Physik, Universit\"at Wien}\\
\small{Boltzmanngasse 5, A-1090 Vienna, Austria}}

\maketitle

\begin{abstract}

Focus is on two parties with Hilbert spaces
of dimension $d$, i.e. ``qudits''.
In the state space of these two possibly entangled qudits an analogue to the
well known tetrahedron with the four qubit
Bell states at the vertices is presented.
The simplex analogue to this magic tetrahedron includes mixed states.
Each of these states appears to each of the two parties as
the maximally mixed state.
Some studies on these states are performed,
and special elements of this set are identified.
A large number of them is included
in the chosen simplex which fits exactly into
conditions needed for teleportation and other applications.
Its rich symmetry -- related to that of a classical phase space -- helps to study entanglement,
to construct witnesses and perform partial transpositions.
This simplex has been explored in details for $d=3$.
In this paper the mathematical background and extensions to arbitrary dimensions
are analysed.
\\[10ex]
\\[3ex]

\end{abstract}

\footnotetext[1]{\texttt{Bernhard.Baumgartner@univie.ac.at}}
\footnotetext[2]{\texttt{Beatrix.Hiesmayr@univie.ac.at}}
\footnotetext[3]{\texttt{Heide.Narnhofer@univie.ac.at}}

\newpage

\section{Introduction}\label{intro}
Entanglement, a non-classical essential feature of quantum theory,
has first been recognized in 1935 in the connection with paradoxes.
Studying it with mathematical accuracy began 1964, introducing the Bell states.
With the new era of proposed applications in teleportation,
quantum computing and quantum communication \cite{BW92},\cite{B93} it became necessary to use
the Bell basis of four ``magic Bell states'' describing the qubit.
Now, in the process of extending the concepts to systems with Hilbert spaces of dimension
greater than two, one faces the practical task of defining analogous sets of states.
The presentation of all possible ideal schemes in \cite{W01} reveals
a large field of structures that can be chosen of.

One item that fits into those schemes is presented in this paper.
We were motivated to study it mainly out of curiosity on the theoretical side.
It has a rich structure of symmetries which enable deep concrete investigations
on the location of the border between entangled and separable states.
(Compare \cite{VW00}.)
It would be no surprise when its mathematical beauty will be reflected
in practical application.

In this paper we focus on two parties with Hilbert spaces of
dimension $d$, i.e. ``qudits''. For these two possibly entangled
qudits we construct an analogue to the well known tetrahedron with
four mutually orthogonal Bell states at the vertices, \cite{HH96}.
This magic tetrahedron includes mixed states, inside lies an
octahedron of separable states. Each of these states appears to each
of the two parties as the maximally mixed state of its qubit.
Considering the duality between maps and states, \cite{ZB04},
these states are related to bistochastic maps, \cite{AU82} (but only for bipartite systems).
We make some remarks on this duality in the concluding Section \ref{summary}.
Now, also for the qudits one might prefer states with this property.
We call it {\bf Locally Maximally Mixed}. For qubits, any such LMM
state can be considered as an element of the tetrahedron,
\cite{BNT02}, but for $d\geq 3$ the analogue statement is no longer
true. So we perform some studies on LMM states in Section \ref{lmm}
and identify special elements of this set. In the following sections
we then recognize a large number of these special states as included
in our chosen subspace of LMM.

For the qudit pair we define and study a special simplex (``generalized
tetrahedron'') $\W$. It has $d^2$ pure states at the vertices, with specified relations between them.
Its rich symmetry helps to study entanglement, and it fits exactly into the
conditions needed for teleportation and dense coding as stated in \cite{W01}.
We explored this magic simplex for $d=3$, \cite {BHN06}.
In this paper we bring a detailed analysis of the mathematical background.
This enable us to
extend the study to higher dimensions.

Choose some basis
$\{|s\rangle \}$ in each factor and define a ``Bell state'', i.e. a maximally entangled pure state, in the Hilbert space
$\C^d \otimes \C^d$ with the vector
\beq \label{omega}
|\Omega _{0,0}\rangle = \sdird \sum_s |s\rangle \otimes |s\rangle .
\eeq
On the first factor in the tensorial product we consider actions
of the \textbf{Weyl operators} defined as
\beqa \label{weyl}
\check{W}_{k,\ell}|s\rangle & = & w^{k(s-\ell)}|s-\ell\rangle ,   \\
w & = & e^{2\pi i/d}, \\
\textnormal {with the identity} \nonumber &&\\
\qquad |s-\ell\rangle & \equiv & |s-\ell +d\rangle .\label{modulo}
\eeqa
The actions of the Weyl operators
produce mutually orthogonal Bell state vectors
\beq
|\Omega_{k,\ell} \rangle = ( \check{W}_{k,\ell}\otimes \one)|\Omega _{0,0}\rangle.
\eeq

The set of index pairs ${(k,\ell )}$ is a
{\bf finite discrete classical phase space}:
$\ell$ denotes the values for the coordinate in ``x-space'', $k$ the values of the ``momentum''.
Remarks on the relation to the physics of the Heisenberg-Weyl quantization we have made
for $d=3$; details on the mathematics follow in Section \ref{groups}.
To each point in this space is associated
the density matrix for the Bell state, the projection operator
\beq \label{proj}
P_{k,\ell} = |\Omega_{k,\ell}\rangle\langle\Omega_{k,\ell}| .
\eeq
The mixtures of these pure states form our object of interest,
the {\bf magic simplex}
\beq
\W\quad =\quad \{ \, \sum c_{k,\ell}P_{k,\ell} \,\, |\,\,\, c_{k,\ell}\geq 0 , \, \sum c_{k,\ell}=1 \, \}\,.
\eeq

As a geometrical object $\W$ is located in a hyperplane of the $d^2$-dimensional
Euclidean space $\{A= \sum a_{k,\ell}P_{k,\ell} \, |\, a_{k,\ell}\in \R\}$
equipped with a distance relation $\sqrt{\Tr (A-B)^2}$.
Specifying the origin $A=0$, it is also equipped
with the Hilbert-Schmidt norm $\sqrt{A^2}$, and the inner product
$\Tr (AB) = \sum a_{k,\ell}b_{k,\ell}$.
All this is imbedded in the $d^4$ dimensional Hilbert Schmidt space of hermitian $d^2\times d^2$ matrices.
We use this Euclidean geometry for ease of calculations.

The main goal in this paper is the exploration of the borders of {\bf SEP},
i.e. the set of separable states. We find that the structure of the
subset SEP$\cap\W$,
the analogue to the octahedron of bipartite qubits, is not quite simple.
It is not a polytope;
but a rather detailed study is enabled by the rich symmetry of the
simplex $\W$.
Using part of it, we determine easily first
two polytopes giving an inner and an outer fence to the border of SEP.
These results, among others, appear in Sections \ref{lmmsets} and \ref{groups}.

Symmetry is then studied in detail in Section \ref{symmetries}.
It simplifies also performing the partial transpositions
of the states in $\W$. This is discussed in Section \ref{partialtrans}. So we get a closer approximation to SEP$ \cap \W$
by studies on {\bf PPT}, that is the set of density matrices remaining positive after partial transposition.
Here we refer to the Peres criterion \cite{P96} which implies that
SEP is a subset of PPT.
Furthermore the partial transposition maps
PPT$\cap\W$ into PPT$\cap \hat \W$,
where $\hat \W$ is another convex subset of LMM, also defined in Section \ref{partialtrans}.
The partial transposition maps SEP onto itself, so the cases of bound entanglement detected
in $\W$ are also cases for bound entanglement in $\hat \W$.

Last, but not least, the symmetry of $\W$ can be exploited as the
symmetry of the set of witnesses \cite{T00} needed there.
Here, in Section \ref{witnesses}, we use the mathematics of convex cones and their duals.
It helps to determine  exactly the borderlines of SEP.
This has been done in \cite{BHN06} for $d=3$.
Extensions to studies for higher dimensions will follow (work in progress).

This study follows two aims.
In the main task of investigations the special simplex is constructed
and its symmetries are stated. Using these symmetries, some details
in the structure concerning entanglement are explored.
In following the second trail we check not to have overlooked anything:
the symmetry group is maximal, the polytopes are optimal.
The proofs of having ``best possible'' results afford some
mathematical subtlety.
We present these subtle investigations in the extra Section \ref{subtleties}.

Various mathematical branches are used: theories of
numbers, groups, convex sets, matrices and Hilbert spaces.
But only some basic facts are needed, to be found in any introduction or encyclopedia, as
\cite{Sch86}, \cite{V64}, \cite{A42}, \cite{W06}.
One side effect, which unfortunately makes some pain, is
the frequent switching of the mathematical points of view. Being too strict
on the reference to the context would make notations cumbersome and difficult to follow.
We try to avoid an overburdening with symbols.
We refer, for example, to the states with the same letters as we use for
the density matrices representing them.
But we are strict on not confusing Hilbert space {\it vectors} with states.
Big Greek letters denote elements of the total Hilbert space,
small letters, mostly in the environment $|\,\,\rangle$, are used for elements of $\C^d$.

Moreover we simplify the notations, omitting the sign for
the tensor product concerning the two parties,
and write e.g. $|\varphi, s\rangle$ for an element of $\C^d \otimes \C^d$
instead of $|\varphi\rangle \otimes |s\rangle$.
Our Unitary Operators $U$, $V$, $W$, that occur in this work acting on the global Hilbert space are all of product form.
They act locally as $\check{U}$, $\check{V}$, $\check{W}$ on the first factor,
and as $\tilde{U}$, $\tilde{V}$, $\tilde{W}$ on the second factor.
$U|\varphi, \psi \rangle$ $=$ $|\check{U}\varphi, \tilde{U}\psi \rangle$.
Mostly $\tilde{U}=\one$, with exceptions in  Section \ref{symmetries}.

\section{LMM states}\label{lmm}

Elements of LMM are the states $\rho$ on the Hilbert space $\Hh_A \otimes \Hh_B$ $=\C^d \otimes \C^d$
which appear locally to each of the single parties as maximally mixed.
\footnote{In the community working with operator algebras,
such a maximally mixed state is known as tracial state.}
The partial trace in one factor gives the maximally mixed state $\omega$
on the other side.

\beq
\rho_A := \Tr_B \rho =\omega_A := \frac 1d \one_A,\quad
\rho_B := \Tr_A \rho =\omega_B := \frac 1d \one_B \label{lmdef}
\eeq

We identify special types of LMM states: Its pure states, isotropic states,
Werner states and maximally exposed elements of SEP$\cap$LMM.

The pure LMM states for qubits are known as the
``Bell states''. We extend this naming to each one of the pure LMM states of qudits.
The single Bell states for fixed $d$ are all unitarily equivalent, involving
local unitary transformations:
Consider the pair $\Omega, \Phi$ of Bell state vectors.
For Schmidt decomposition, we choose a preferred  basis $\{|s\rangle \}$ in $\Hh_B$.
Then there are two different bases $|\psi_s\rangle$ and  $|\varphi_s\rangle$ in $\Hh_A$,
such that
\beq
|\Omega\rangle =\sdird \sum_s |\psi_s,s\rangle,
\qquad |\Phi\rangle =\sdird \sum_s |\varphi_s,s\rangle.
\eeq
$|\Omega\rangle$ is mapped to $|\Phi\rangle$ by extension of the local unitary operator
\beq
\check{U}=\sum_s |\varphi_s\rangle\langle\psi_s|. \label{upair}
\eeq
Mixtures of a Bell state $\Omega$ with the global maximally mixed state
\beq
\omega=\frac{1}{d^2}\one
\eeq
define the {\bf isotropic states} $(1-\alpha )\omega +\alpha |\Omega\rangle\langle\Omega|$.
Again all the isotropic states with the same $\alpha$ are unitarily equivalent.

Other special LMM states are the lines of {\it Werner states}, related to the lines of isotropic states
by {\bf PT}, that is {\bf Partial Transposition}. See \cite{VW00} for the appropriate
ranges of the parameters $\alpha$ and other details.
They are also all equivalent. We need no special check of their belonging to LMM.
There is the general fact:
\begin{lem}\label{lmmpt}
The LMM property (\ref{lmdef}) is preserved under partial transposition.
\end{lem}
\begin{proof}
The equation (\ref{lmdef}) is equivalent to the statement that for each $\varphi \in \C^d$
and each  basis $\{\psi_t\}$
\beq \label{lmdef2}
\sum_t \langle\varphi,\psi_t|\rho|\varphi,\psi_t\rangle = \|\varphi\|^2/d.
\eeq
We write this as $\Tr \rho\, Q = \|\varphi\|^2/d$ with $Q:= \sum_t |\varphi,\psi_t\rangle \langle\varphi,\psi_t|$.
The PT operator in the Hilbert Schmidt space is symmetric, i.e.
$\Tr (PT(\rho)\cdot Q)=\Tr(\rho \cdot PT(Q))$.
Calculation of PT is done in the preferred basis $|s\rangle$;
the expansion $|\varphi\rangle =\sum_s \varphi_s |s\rangle$ gives
\beqa
PT(Q)=\sum_{s,r,t}\varphi_s \varphi_r^\ast PT(|s,\psi_t\rangle \langle r,\psi_t|)\nonumber\\
=\sum_{s,r,t}\varphi_s \varphi_r^\ast |r,\psi_t\rangle \langle s,\psi_t|=
\sum_t |\hat\varphi,\psi_t\rangle\langle\hat \varphi,\psi_t| ,
\eeqa
with $|\hat\varphi\rangle=\sum_r\varphi_r^\ast|r\rangle$.
The complex conjugation does not change the norm
and (\ref{lmdef2}) holds for PT$(\rho)$.
\end{proof}
The Werner states do not only have special symmetries, they have the property of attaining
the minimal possible distance (see \cite{GB02}) to the maximally mixed state $\omega$
when they are at the border between SEP and the entangled states.
We conjecture that they are the only LMM states with this property. What we can show easily is the
\begin{lem}
If an LMM state at the border between PPT and non-PPT states has the
minimal border distance to $\omega$, which is $1/d\sqrt{d^2-1}$,
then it is a Werner state.
\end{lem}
\begin{proof}
Performing PT on the density matrix $\rho$ of this state we get a density matrix
$\sigma$ at the border of PPT$\cap$LMM to non-positive matrices.
The matrices with minimal distance at that border have the form $\sigma=(\one -P)/(d^2-1)$,
with $P$ a projector belonging to a pure state.
Pure LMM states are Bell states, so $\sigma$ is isotropic
and $\rho=PT(\sigma)$ is a Werner state.
The Euclidean distance squared is easily calculated as
\beq
\Tr (\rho -\omega)^2=\Tr (\sigma-\omega)^2=\Tr\sigma^2 -\frac1{d^2}=\frac1{d^2(d^2-1)}.
\eeq
\end{proof}
$\W$ does not contain Werner states  if $d\geq 3$, but
$\hat\W$ does; see Section \ref{partialtrans}.

A third kind of special LMM-states appears in both $\W$ and $\hat\W$:
The separable states with the largest possible distance between $\omega$
and SEP$\cap$LMM.
\begin{thm}\label{maxd}
The maximal distance of a $\sigma\in{\rm SEP}\cap$LMM to $\omega$ is $\sqrt{d-1}/d$.
It is attained if and only if the density matrix $\sigma$ has the form
\beq
\sigma=\frac1d\sum_s|\varphi_s, \psi_s\rangle\langle \varphi_s, \psi_s|,
\eeq
where both $\varphi_s$ and $\psi_s$ are bases for $\C^d$.
\end{thm}
\begin{proof}
The first condition, $\sigma \in$SEP, is fulfilled iff $\sigma$ can be represented as
\beq\label{sepcond}
\sum_\alpha \lambda_\alpha|\varphi_\alpha, \psi_\alpha\rangle\langle \varphi_\alpha, \psi_\alpha|
\eeq
with $\lambda_\alpha> 0$ , $\sum_\alpha \lambda_\alpha=1$, and normed vectors
$\varphi_\alpha$, $\psi_\alpha$.\\
The second condition, $\sigma \in$LMM, implies that
$\forall\varphi, \forall\psi$ with norm one
\beq\label{lmmcond}
\langle\varphi,\psi|\sigma|\varphi,\psi\rangle \leq
\sum_j \langle\varphi,\psi_j|\sigma|\varphi,\psi_j\rangle
=\langle\varphi|\sigma_A|\varphi\rangle
=\frac1d,
\eeq
where we considered some basis $\psi_j$ containing the given $\psi$.
Applying (\ref{lmmcond}) to the vectors appearing in (\ref{sepcond})
gives
\beq\label{comb}
\Tr \sigma^2=
\sum_\alpha \lambda_\alpha\langle\varphi_\alpha, \psi_\alpha |\sigma|\varphi_\alpha, \psi_\alpha\rangle
\leq\sum_\alpha \lambda_\alpha\frac1d =\frac1d.
\eeq
This proves the first statement about the maximal distance, since
$\|\sigma-\omega\|^2=\Tr\sigma^2-1/d^2$.
To prove the second statement observe that the inequality (\ref{lmmcond})
turns to an equality iff $\forall\psi^\bot$
with $\langle\psi^\bot|\psi\rangle=0$
the equality $\langle\varphi,\psi^\bot|\sigma|\varphi,\psi^\bot\rangle=0$ holds.
The same is true with the roles of the two sides interchanged, that is
$\forall\varphi^\bot$ with
$\langle\varphi^\bot|\varphi\rangle=0$
one has $\langle\varphi^\bot,\psi|\sigma|\varphi^\bot,\psi\rangle=0$.
So one can start diagonalizing the matrix $\sigma$.
One begins with one pair of vectors appearing in(\ref{sepcond}), say $\alpha=0$.
\beq
\sigma =\frac1d|\varphi_0,\psi_0\rangle\langle\varphi_0,\psi_0|+\frac{d-1}d \sigma_{d-1}.
\eeq
The matrix $\sigma_{d-1}$ is a normalized density matrix in the LMM$\cap$SEP
with {\it lower dimension}, acting on $\C^{d-1}\otimes\C^{d-1}$.
This can easily be seen by
$(d-1)\Tr_B \sigma_{d-1}=d\Tr_B \sigma -|\varphi_0\rangle\langle\varphi_0|
=\one_A-|\varphi_0\rangle\langle\varphi_0| = (d-1) \one_{A,d-1}$
and
$\langle\varphi^\bot,\psi^\bot|\sigma_{d-1}|\varphi^\bot,\psi^\bot\rangle
=d/(d-1)\langle\varphi^\bot,\psi^\bot|\sigma|\varphi^\bot,\psi^\bot\rangle$.
Now one may proceed inductively, expanding $\sigma_{d-1}$ in the form (\ref{sepcond})
-- generally with new vectors -- diagonalizing
$(d-1)\sigma_{d-1} =|\varphi_1,\psi_1\rangle\langle\varphi_1,\psi_1|+(d-2) \sigma_{d-2}$,
and so on.
\end{proof}

Each one of these maximally exposed SEP$\cap$LMM states
is in unique correspondence to a pair of bases in the Hilbert spaces of the parties
and a one two one mapping between them. Each one can be represented as a mixture of $d$ Bell states
appearing in some $\W$. For example one may use the bases
characterizing $\sigma$ as stated in Theorem \ref{maxd}, to construct a simplex $\W$:
Put them into the definition (\ref{omega}),
$|\Omega _{0,0}\rangle = \sdird \sum_s |\varphi_s ,\psi_s\rangle $,
construct the $P_{k,0}$, and represent $\sigma=\sum_k P_{k,0}/d$.
But this representation is not unique.
More about this is presented in Section \ref{groups}.
PT maps this set of maximally exposed states onto itself, so these states appear in $\hat\W$ also.

\section{Subsets of LMM}\label{lmmsets}

Let us proceede and look at {\it subspaces} of LMM.
Most important are the Bell states appearing in the chosen subspace.
{\it Any} set of $d^2$ mutually orthogonal Bell states $P_\alpha$ --
orthogonality of the Hilbert space vectors
$\langle\Omega_\alpha|\Omega_\beta\rangle = \delta_{\alpha,\beta}$
is in this case equivalent to the orthogonality of the density matrices in the
Euclidean space $\Tr P_\alpha P_\beta=\delta_{\alpha,\beta}$
 -- span a maximal simplex.
Each Bell state comes with an optimal witness, a hyperplane $B_\alpha$
defined as $B_\alpha:=\{\rho\,:\, \Tr \rho (P_\alpha -\one /d)=0\}$.
These $d^2$ hyperplanes $B_\alpha$, together with the $d^2$ hyperplanes
$A_\alpha:=\{\rho\,:\, \Tr \rho P_\alpha=0\}$ containing the faces of the simplex,
define an {\bf enclosure polytope}. Outside of it are only entangled states.
The projectors $P_\alpha$ generate a maximal abelian subalgebra
of operators acting on $\Hh$.
So these conditions alone bring already some insight, but they still
allow for many different choices of an LMM subspace.
They are not all equivalent.
The geometric symmetry of the enclosure polytope, the same symmetry as that of the simplex,
is deceptive: SEP must be inside, but the relations of its detailed geometry
to the set of pure states in the chosen subspace depends on their algebraic relations.
The single Bell states are equivalent,
but already {\it pairs} of orthogonal Bell states fall into different classes of pairs if $d\geq 4$.
Enter spectral theory:
Each class is characterized by the spectrum of the local unitary operators
identified in (\ref{upair}) connecting the pair.
Orthogonality of the Bell state vectors implies $\Tr U =0$.
Being interested only in the {\bf intertwining relation}
\beq\label{intertw}
U|\Omega\rangle\langle\Omega|U^\dag =|\Phi\rangle\langle\Phi|
\eeq
we are free to choose a phase factor for $U$ such that one of its eigenvalues is equal to $1$.
The condition $\Tr U=0$ specifies the rest of the spectrum only for qubits and qutrits.
For $d\geq 4$ there are various possibilities,
defining different classes of equivalent pairs:
Unitary or antiunitary local mappings of one pair onto the other
can be applied to the interwiners $U$. So their spectra are either
unchanged or complex conjugated and rotated. This characterizes the classes.

To choose {\it special sets} of Bell states an extra criterion which a theoretician
likes to pose is that the intertwining operators form a unitary group,
allowing for multiplication of any two of them.
This gives a strong restriction on their spectra. Enter number theory:
\begin{thm}\label{intergroup}
If $\{U^n\}$ is a group of intertwiners between mutually orthogonal Bell states, then,
with an appropriately chosen overall phase factor,
$U$ has eigenvalues $e^{2\pi\, i\, m/b}$, where $0\leq m\leq b-1$,
and $b$ is either a divisor of $d$ or equal to $d$.
Considering intertwiners acting as $U =\check{U}\otimes \one$ on $\Hh_A$ only, the multiplicity
of each eigenvalue of $\check{U}$ is $d/b$.
\end{thm}
\begin{proof}
The Euclidean space of density matrices has finite dimension,
the set of orthogonal projectors onto $U^n|\Omega\rangle$ is finite, less than $d^2$,
and there exists some smallest natural number $b$, such that $U^b=\one\cdot \,\textrm{{\it phasefactor}}$.
We choose the phasefactor for $U$ in such a way that we have $U^b=\one$.
In the following we consider $U$ acting on $\Hh_A$ only.
Since $\Tr U^b=d\, \Tr\check{ U}^b$, the orthogonality of the Bell states implies,
as stated before equ. (\ref{intertw})
\beq\label{cycl}
\Tr \check{U}^n=d\,\delta_{n,0} \quad {\rm for} \quad 0\leq n\leq b-1 .
\eeq
$\check{U}^b=\one$ implies that the eigenvalues of $\check{U}$ are elements of
$\{e^{2\pi\, i\, m/b},\,\, 0\leq m\leq b-1\}$
Denote the multiplicities as $f(m)$.
Then the equation (\ref{cycl}) can be read as a formula for the
Fourier transform of $f(m)$. The inverse transform gives
\beq
f(m)=\frac1b \sum_{n=0}^{b-1} e^{2\pi\, i\, m\,n/b}\,\Tr \check{U}^n =\frac{d}{b}.
\eeq
This number has to be an integer.
\end{proof}
\begin{cl}
Any group of unitary intertwiners between mutually orthogonal Bell states contains
finite cyclic subgroups. Each one is of some order $b$, where either $b=d$, or $b$
is a divisor of $d$.
\end{cl}
It follows that there are not many different possibilities for structures of such groups.
Our choice is possible for all $d$, whether prime or not.

\section{Groups and the classical phase space for the magic simplex} \label{groups}

Letters of the set $\{j,\ldots t\}$
denote numbers $0, 1 \ldots d-1$. They are considered as elements of $\Z_d := \Z/d\Z$.
Calculations with them are to be understood as ``modulo d''.

{\bf Intertwiners} are the Weyl operators $W_{k,\ell} =\check{W}_{k,\ell}\otimes \one$ presented in Section \ref{intro}:
\beq \label{weylaction}
W_{k,\ell} P_{p,q} W_{k,\ell}^\dag = P_{p+k,q+\ell}\,.
\eeq
The Weyl operators obey the {\bf Weyl relations}
\beqa \label{Weylrel}
W_{j,\ell}W_{k,m} & = & w^{k\ell}W_{j+k,\ell +m} \, ,         \\
W_{k,\ell}^\dag  = W_{k,\ell}^{-1} & = & w^{k\ell}W_{-k,-\ell} \, ,       \\
W_{0,0} & = & \one .
\eeqa
They form the Heisenberg-Weyl group $\Ww$.
More precise: $\Ww$ is a finite discrete subgroup of the
doubly infinite continuous Heisenberg group; compare \cite{W06}.
Group elements are $w^m W_{k,\ell}$. The phase factors $\{w^m\one\}$ form an abelian normalizer;
the factor group is $\Ww /\Z_d \cong \Z_d \times Z_d$.
This can be considered in the sense originally meant by Weyl, \cite{W31}, as the
quantization of classical kinematics.
The kinematics of the Galilei group is represented in the discrete classical phase space
as $\Z_d\times\Z_d$, generated by the global boost $(p,q)\mapsto(p+1,q)$
and the global space translation $(p,q)\mapsto(p,q+1)$.

The classical phase space
$\T :=\{(p,q)\}$ is a lattice on a two-dimensional torus.
It has a ``linear'' structure -- multiplication
by constants and addition is always done in the ring $\Z_d$ -- and it
is a symmetric space for the Heisenberg-Weyl group:
We define the action of $\Ww$ on $\T$ by identifying each phase space point
$(p,q)$ with the projector $P_{p,q}$ and use equ. (\ref{weylaction}).
Moreover we identify non-negative normalized densities $\{c_{p,q}\geq 0\, ,\,\sum c_{p,q}=1\}$
with the elements $\sum c_{p,q}P_{p,q}$ of $\W$.
Special use is made of equidistributions over subsets $Q\subset \T$
and the corresponding density matrices
\beq
\rho_Q:=\sum_{(p,q)\in Q} P_{p,q}/|Q|.
\eeq
Now the group structure of $\Ww$ gives a first insight into the structure of SEP$\cap\W$.
Each cyclic subgroup $\{W_{k,\ell}^n\}$ acting on a point $(p,q)$ of $\T$ generates a
{\bf line} $\{(p+nk,q+n\ell )\}$.
\footnote{Warning: These lines do not for each $d$ fulfill the conditions for a ``line'' in
the sense of affine geometry. See also \cite{B04}.}
These lines, for $d$ prime, have been identified in \cite{N06} as corresponding to separable states.
If there are non-cyclic abelian subgroups -- which may be the case if $d$ is not prime --
they generate sublattices, each one with
at most two independent basis vectors.
\begin{proposition}\label{seprho}
Each line or sublattice with d points is generated by an abelian
subgroup of $\Ww$ and corresponds to a maximally exposed state in SEP$\cap\W$.
\end{proposition}
\begin{proof}
Consider a sublattice $Q$ with $d$ points. A lattice in $2$ dimensions can be
represented with $2$ basis vectors.
So we can represent
\beq
Q=\{(b+j\mu+k\nu, q+\ell\mu+m\nu),\, 0\leq\mu\leq b-1, 0\leq\nu\leq c-1, b\cdot c=d\}
\eeq
We include the cases $b=d$, $c=1$, representing lines.
For the matrices the representation is
\beq\label{rhoqu}
\rho_Q=\frac1d \sum_\mu\sum_\nu U^\mu V^\nu |\Omega_{p,q}\rangle\langle\Omega_{p,q}|U^{-\mu} V^{-\nu}
\eeq
where $U=e^{i\gamma} W_{j,\ell}$, $V=e^{i\delta} W_{k,m}$,
with the phase factors chosen, if necessary, such that $U^b=V^c=\,\one$.
Each one of the smaller exponents gives other elements of $\Ww$; so
\beq\label{deltas}
\Tr_A U^\mu V^\nu = \delta_{\mu,0}\delta_{\nu,0}\cdot d
\eeq
For the sublattice the Weyl relations (\ref{Weylrel}) imply
\beq\label{permute}
U\cdot V\cdot U^\dag\cdot V^\dag=w^{k\ell-jm}\one ,
\eeq
and the exponent $k\ell-jm$ is the oriented area of a unit cell of $Q$.
The union of all $d$ cells spans all of $\T$, which has area $d^2$,
once or several times; so
$(k\ell-jm)\cdot d =z\cdot d^2$, with some $z\in\Z$.
It follows that $k\ell-jm\equiv 0$, the r.h.s. of (\ref{permute})is $\one$,
so $U$ and $V$ commute.
This allows for a common spectral decomposition
\beqa\label{spec}
\check{U}&=&\sum_{s=0}^{d/b-1}\sum_{t=0}^{d/c-1}f(s,t)e^{2\pi i\,s/b} |\varphi_{s,t}\rangle\langle\varphi_{s,t}|,\nonumber\\
\check{V}&=&\sum_{s=0}^{d/b-1}\sum_{t=0}^{d/c-1}f(s,t)e^{2\pi i\,t/c} |\varphi_{s,t}\rangle\langle\varphi_{s,t}|.
\eeqa
We get the multiplicity function $f$, in a way analogous to the proof of Theorem  \ref{intergroup}.
Here we use the Fourier transform in two variables, and equ.(\ref{deltas}):
\beq
f(s,t)=\frac1{b\cdot c} \sum_{\mu,\nu}e^{2\pi i\,(\mu s/b +\nu t/c)}\Tr \check{U}^\mu \check{V}^\nu=1.
\eeq
The diagonalizing basis $\varphi_{s,t}$ in $\Hh_A$ is now used for a
Schmidt decomposition of the Bell state vector.
\beq\label{oschmidt}
|\Omega_{p,q}\rangle = \sdird \sum_{s,t}|\varphi_{s,t},\psi_{s,t}\rangle,
\eeq
With $\psi_{s,t}$ as the appropriate basis in $\Hh_B$.
Inserting (\ref{spec}) to (\ref{oschmidt}) into  (\ref{rhoqu}),
the summation over the phase factors brings some $\delta$ factors,
reducing the summations.
The result is
\beq
\rho_Q = \frac1d \sum_{s,t}|\varphi_{s,t},\psi_{s,t}\rangle\langle\varphi_{s,t},\psi_{s,t}|.
\eeq
This expression for $\rho_Q$ is exactly as it is used for the matrices in Theorem \ref{maxd}.
\end{proof}

SEP is a convex set and the separable states $\rho_Q$ identified in Proposition \ref{seprho}
can be considered as the extreme points of a {\bf kernel polytope}
which is a subset of SEP$\cap$LMM.
These $\rho_Q$ appear also as extreme points of the enclosure polytope,
but do not cover all of them if $d\geq 3$.
For the vertices of the enclosure polytope the set $Q$ can be any subset with $d$ elements
of $\T$; for the kernel polytope this set $Q$ has to be a line or a sublattice.
That all the other sets $Q$ correspond {\it in fact} to entangled states is proven in
the Theorem \ref{seprhothm} in Section \ref{subtleties}.
\begin{thm}\label{numberofextrems}
The number of lines and sublattices with $d$ points in $\T$ is
$$N(d)=d\cdot \left[ 1+d+\sum b \,\right] ,$$
where the sum runs over all the $b$ which are proper divisors of $d$.
\end{thm}
\begin{proof}
The number in square brackets must be the number of lines and lattices $Q$ of order $d$, each containing
the point $(0,0)\in\T$. All the others can be found by translations; and
doing all $d^2$ translations gives each line and each lattice of order $d$ in $d$-fold multiplicity.

We give a list of these $Q$ containing $(0,0)$:
\begin{description}
\item[a)]\quad $\{(s,0)|\quad 0\leq s\leq d-1\}$,\quad\quad one line
\item[b)]\quad $\{(k\cdot s,s)|\quad 0\leq s\leq d-1\}$,\quad d lines, one for each $k\in[0,\ldots d-1]$
\item[c)]\quad $\{(\mu\cdot b+\nu\cdot b,\,\nu\cdot d/b)|\quad
0\leq \mu\leq d/b-1,\,0\leq \nu\leq b-1\}$,\quad b sublattices, one for each $\nu\in[0,b-1]$,
with $b$ a proper divisor of $d$.
\end{description}
\end{proof}
We remark that some of the sublattices listed in {\bf c)} can also be
considered as lines. But not all of them, if $d$ is not a simple product
of prime numbers but contains also squares or higher powers of them.
One example is $d=4$, $b=2$, $\nu=0$.

\section{Symmetries of $\W$}\label{symmetries}

We are looking for symmetries compatible with the entanglement,
just to make the investigations simpler.
We do not pose detailed restrictions, no measure for entanglement is needed.
Just the following, physically motivated characterization is sufficient:
\begin{define}\label{ecompatible}
A mapping $L:\,\W \rightarrow\W$ is E-compatible (i.e. compatible with entanglement), iff
\begin{description}
\item[a)]\quad Bell states are mapped to Bell states,
\item[b)]\quad $L$ is mixture preserving
$$L(\alpha\rho+(1-\alpha)\sigma)=\alpha L(\rho)+(1-\alpha)L(\sigma),$$
\item[c)]\quad SEP$\cap\W$ is mapped onto itself.
\end{description}
\end{define}
Now if we have a {\it local} unitary transformation $\rho\rightarrow U\rho U^\dag$,
the separability is preserved.
Also the conditions \textbf{a)} and \textbf{b)} are fulfilled.
So we know already about a subgroup of symmetry transformations:
The translations of phase space; they are implemented as local unitary transformations
in the Heisenberg Weyl group $\Ww$.

For general $L$ the fulfilling of \textbf{b)}
implies the possibility to linearly extend $L$.
It gives then, due to \textbf{a)}, an invertible norm preserving
linear map of the Euclidean space, spanned by the $P_{k,\ell}$, onto itself.
It effects a permutation of the Euclidean basis elements $P_{k,\ell}$,
hence a map $\T\rightarrow\T$. Vice versa, any permutation of this kind
extends via mixing preserving to a map $\W\rightarrow \W$.
Now, after any permutation of $\T$ a certain translation can bring the origin $(0,0)$
back to its place. So each of the symmetry operations
can be formed as a product of a phase space translation with a certain point transformation $M$,
which leaves the point $(0,0)$ at its place.

The {\it tool box} of these point transformations of phase space contains:
\begin{description}
\item[The ``horizontal'' shear of phase space], $\mathcal{H}:$
$\qquad
\left( \begin{array}{c} p \\ q \end{array} \right) \mapsto
\left( \begin{array}{c} p \\ p+q \end{array} \right).
$
Its powers form a cyclic subgroup.
The elements are represented by the matrices $ \left( \begin{array}{cc} 1 & 0 \\ n & 1 \end{array} \right)$.
(In discrete classical mechanics this is a free time evolution.)
\item[The ``vertical'' shear of phase space], $\mathcal{V}:$
$\qquad
\left( \begin{array}{c} p \\ q \end{array} \right) \mapsto
\left( \begin{array}{c} p+q \\ q \end{array} \right)$.
The elements of the generated cyclic subgroup are represented by the matrices
$ \left( \begin{array}{cc} 1 & n \\ 0 & 1 \end{array} \right)$.
(It may be considered as a local boost.)
\item[A quarter rotation of phase space], $\mathcal{R}:$
$\qquad
\left( \begin{array}{c} p \\ q \end{array} \right) \mapsto
\left( \begin{array}{c} q \\ -p \end{array} \right)$.
It is represented by the matrix
$ \left( \begin{array}{cc} 0 & 1 \\ -1 & 0 \end{array} \right)$.
$\qquad\mathcal{R}^2 =-\one$ is a point reflection.
\item[Squeezing], a scale transformation
$\qquad
\left( \begin{array}{c} p \\ q \end{array} \right) \mapsto
\left( \begin{array}{c} r\cdot p \\ s\cdot q \end{array} \right)$,
for $r\cdot s \equiv d+1$, possible for each $r$ relative prime to $d$.
\item[Reflections]:

\begin{itemize}
\item
Inversion of momentum, $\mathcal{S}:$ $p\mapsto -p$
\item
Space reflection,  $q\mapsto -q$
\item
Diagonal reflection,  $p\mapsto q,\quad q\mapsto p$
\end{itemize}

\end{description}

\begin{proposition}\label{esymplectic}
Consider a linear mapping $\T \rightarrow \T$, defined as the application of a
$2\times 2$ matrix $M$ with elements $\in \Z_d$,
and with $\,\det M =\pm 1$.
By extending it to a mapping $\W \rightarrow \W$, it is E-compatible.

These matrices form the extended symplectic group $\textrm{Sp}(2,\Z_d)$.
\end{proposition}
\begin{proof}
Addition and multiplication of the matrix elements is according
to the rules of the ring $\Z_d$. This defines the matrix multiplication.
The unit matrix has $\det\one=1$ and is also in this set. Inverting
a general element $M$ is achieved with the mapping
\beq\label{mmatrix}
M=\left( \begin{array}{cc} k & m \\ \ell & n \end{array} \right) \quad\mapsto\quad
M^{-1}=\pm\left( \begin{array}{cc} n & -m \\ -\ell & k \end{array} \right),
\eeq
with the sign equal to the sign of $\det M$.
So the matrices which are either \textit{symplectic}, $\det M=1$, or
\textit{mirror symplectic}, $\det M=-1$, form a group.
We establish now the three transformations $\mathcal{V},\,\mathcal{R},\,\mathcal{S}$
as generating elements:

First, they generate $\mathcal{H}=\mathcal{R}^{-1}\mathcal{V}\mathcal{R}$
and all the powers $\mathcal{V}^t ,\mathcal{H}^t$.
The space reflection is $\mathcal{R}^{-1}\mathcal{S}\mathcal{R}$,
diagonal reflection $\mathcal{R}\mathcal{S}=-\mathcal{R}^{-1}\mathcal{S}$,
and squeezing is $\mathcal{R}\mathcal{V}^{-s}\mathcal{H}^r \mathcal{V}^s$.
Multiplication by $\mathcal{S}$ maps symplectic to mirror symplectic matrices
and vice versa. Consider a general symplectic $M$.
The condition $\det M=1$ can only then be true if $k$ and $m$
are relative prime. Also $k$ and $\ell$ have no common divisor.
(Letters denoting the matrix elements are placed as in (\ref{mmatrix}).)
So there exist a $t$ and an $s$, such that $m+k\,t\equiv 0$
and $\ell+k\,s\equiv 0$.
One calculates $$M\mathcal{H}^t =\left( \begin{array}{cc} k & 0 \\ \ell & n+t\,\ell \end{array} \right),
\qquad \mathcal{H}^s M\mathcal{H}^t =\left( \begin{array}{cc} k & 0 \\ 0 & n+t\,\ell \end{array} \right).$$
This matrix performs a squeezing. It can be represented as stated above.
Multiplication by $\mathcal{H}^{-s}$ from the left and $\mathcal{H}^{-t}$ from the right
gives us back the matrix $M$.

To realize the E-compatibility we proceed as we did for $d=3$.
A general group element $M$ can be considered as a product of
the three generating elements. For these we present the operators
$C,U_\mathcal{R},U_\mathcal{V}$ and use their products as $U_M$.
The E-compatibility of the generating elements infers so the E-compatibility of $M$.

Now we construct for each generating group element $M$ an
operator $\check{U}_M\otimes \tilde{U}_M^{-1}$, either unitary or anti-unitary.
Its local action in the factor $\C^d$ on the left hand side as $\check{U}_M$
transforms the Weyl operators, unaffected by $\tilde{U}_M$, as
\beq\label{phaseweyl}
U_M W_{k,\ell}\,U_M^{-1}= e^{i\eta(M,k,\ell)}W_{k',\ell '},
\eeq
when $M$ maps $(k,\ell)\mapsto (k',\ell ')$.
Some phase factors $e^{i\eta}$ may appear.
Then we use operators $\tilde{U}$ acting in $\C^d$ on the right hand side.
They are uniquely defined by the condition that
$\check{U}\otimes \tilde{U}^{-1}$
leaves the chosen Bell state vector $\Omega_{0,0}$ invariant.
Its matrix elements in our preferred basis are
$\langle s|\tilde{U}|t\rangle =\langle t|\check{U}|s\rangle$.
The joint action in the space $\W$ can now be calculated to give
\beq\label{waction}
P_{k,\ell}\quad\mapsto\quad (\check{U}_M\otimes \tilde{U}_M^{-1})P_{k,\ell}(\check{U}_M^{-1}\otimes \tilde{U}_M)=
U_M P_{k,\ell }U_M^{-1}=P_{k',\ell '}.
\eeq

Now we look at an implementation of the generating elements
as local transformations of the Hilbert space.
The reflection $\mathcal{S}$ can be implemented by complex conjugation in the preferred basis
$$\check{C}:\quad \sum_s\varphi_s|s\rangle \mapsto \sum_s\varphi_s^\ast|s\rangle.$$
This is a local anti-unitary operation. It acts onto the Weyl operators as
\beq\label{reflectionlocal}
\check{C}\,\check{W}_{k,\ell}\,\check{C}= \check{W}_{-k,\ell}.
\eeq
Its anti-linear extension $C$ is complex conjugation in the global Hilbert space,
\beq\label{reflection}
C\,W_{k,\ell}\,C= W_{-k,\ell},
\eeq
mapping SEP onto SEP. So $\mathcal{S}$ is E-compatible.

The other two generators are implemented by local unitaries, so
the E-compatibility is obvious.
The quarter rotation $\mathcal{R}$ is implemented
as  $U_\mathcal{R}=\check{U}_{\mathcal{R}}\otimes \tilde{U}_\mathcal{R}^{-1}$ by the local Fourier transform:
$$\check{U}_{\mathcal{R}}:\quad |s\rangle \mapsto \sdird \sum_t w^{-s\, t}|t\rangle .$$
It acts onto the Weyl operators as:
\beq\label{rotation}
U_{\mathcal{R}}W_{k,\ell}U_{\mathcal{R}}^\dag= w^{-k\, \ell}\,W_{\ell ,k}.
\eeq
For implementing the vertical shear of phase space, $\mathcal{V}$,
one may choose any integer $\nu$ and define
$$\check{U}_{\mathcal{V}}:\quad |s\rangle \mapsto w^{-s(s+d+2\nu )/2}|s\rangle.$$
For general dimension $d$ we use the ordinary integers $s\in \{0,1\dots , d-1\}$
when calculating the exponents.
(In $\Z_d$ dividing by $2$ is well defined for odd $d$ only.)
For even $d$ the half-integer powers of $w$ have to be chosen consistently for all the odd $s$.
This choice, e.g. $w^{\mu /2}=e^{i\pi\mu /d}$, appears then also in
the action onto the Weyl operators:
\beq\label{shear}
U_{\mathcal{V}}W_{k,\ell}\,U_{\mathcal{V}}^\dag= w^{\ell(\ell +d+2\nu)/2}\,W_{k+\ell,\ell}.
\eeq
\end{proof}

Now this group of E-compatible point transformations is maximal,
other transformations do not have the compatibility property.
The proof is given in Section \ref{subtleties}.

Some remarks on more subtleties:
Note that the implementations of the matrix group elements $M$ are not unique.
There are four different groups of transformations of vectors and operators involved.
Transformations of vectors in $\C^d$ by $\check{U}_M$, unitary transformations of the operators
as noted in (\ref{phaseweyl}). \footnote{Some of the unitary transformations of operators appear as ``gates'' in Quantum
Computation, see e.g. \cite{G99}, \cite{GKP01}.}
Then there are transformations by $\check{U}_M\otimes\tilde{U}_M$ of vectors in the global Hilbert space,
and the related transformations of the operators
as noted in (\ref{waction}).
Only the last one gives a representation of Sp$(2,\Z_d)$, when
restricted to the Euclidean space spanned by the $P_{k,\ell}$.
The others are ``quantizations'', involving phase factors.
There is moreover the possibility to multiply each $U_M$ by some $W_{j,r}$,
and this gives a discrete set of different implications.

The total group $\mathbb{E}$ of E-compatible transformations has, by the way, the structure of the
semidirect product $\textrm{Sp}(2,\Z_d)\rtimes\Ww$
of the extended symplectic group and the Heisenberg-Weyl group.
That is (see, for example \cite{W06}):
The extended symplectic group gives an automorphism of
the normal subgroup $\Ww$, as noted in (\ref{phaseweyl}).
The group product in $\mathbb{E}$ is given as
\beq
(U_M W_{j,m})*(U_L W_{k,\ell})=(U_{M\cdot L}\,e^{i\eta(L,j,m)}W_{j'+k,\,m'+\ell})
\eeq
when $L$ maps $(j,m)\mapsto (j',m ')$.

\section{Witnesses and more symmetry}\label{witnesses}

Entanglement witnesses have been introduced, \cite{T00}, to detect the entanglement.
Detection can be either experimentally or theoretically.
To prove Theorem \ref{seprhothm} in Section \ref{subtleties}
we use them in that way, to discern the entangled vertices of the enclosure polytope
from the separable ones.

Here we reverse the point of view. ``Witnesses'' are used to study the location of SEP,
the convex set of separable states. We define the set of {\it structural} witnesses,
\beq
{\rm SW}:= \{K=K^{\dag}\neq 0 \,\,|\,\, \forall \sigma \in \textrm{SEP}:\quad \Tr (\sigma K)\geq 0 \}\, .
\eeq
This set forms a convex cone of operators. SW$\cup \{0\}$ is the dual convex cone to
$\{\alpha\rho,\, \alpha\geq 0,\,\rho\in \textrm{SEP}\}$
and thus completely characterizes the location of SEP.
Geometrically, every structural witness defines a hyperplane in the Hilbert-Schmidt space
of hermitian matrices $\rho$, which is a Euclidean space with dimension $d^4$.
The extremal rays of this dual cone are {\it tangential} witnesses for
density matrices $\rho$ on the {\it surface} of SEP.
\beq
{\rm TW}:= \bigcup_{\rho\in \, {\it surface}\,({\rm SEP})} {\rm TW}_\rho
\eeq
\beq
\rho\in\,{\it surface}\,({\rm SEP}):\qquad {\rm TW}_\rho := \{K\in {\rm SW} \,|\quad  \Tr (\rho K)= 0\, \} .
\eeq

Being interested in SEP restricted to a linear subspace of states, we may restrict the study of witnesses
onto a dual subspace. If the set of states  is defined by invariance
under the action of a group $\G$,
the dual subspace is a set of witnesses which are also  invariant.
The details of this argument have been presented for $d=3$.
There was no use of a special dimension
and we may take over the results from \cite{BHN06}:
\begin{thm}\label{charwitness}
Characterizing SEP$\cap\W$ through witnesses is simplified by using the following properties:
\begin{itemize}
\item SEP$\cap\W$ is completely characterized by duality, using
witnesses of the form $K=\sum_{k,\ell}\kappa_{k,\ell}P_{k,\ell}$.
\item Such an operator $K$ is a witness, iff \,$\forall \tilde{\psi} \in \C^d$ the operator
\beq
\sum_{k,\ell}\kappa_{k,\ell}W_{k,\ell}|\tilde{\psi}\rangle\langle\tilde{\psi}|W_{k,\ell}^{-1}
\eeq
is not negative.
\item $K$ is a tangential witness in some TW$_\rho$ iff \,$\,\exists |\varphi ,\psi\rangle$
such that
\beq
\sum_{k,\ell}\kappa_{k,\ell}\,|\langle\varphi|W_{k,\ell}|\tilde{\psi}\rangle|^2=0
\eeq
with $|\tilde{\psi}\rangle=\sum_s \langle s|\psi\rangle^\ast |s\rangle$.
The state
\beq
\rho=\langle \, |\varphi ,\psi\rangle\langle\varphi ,\psi| \, \rangle_\G
\eeq
is a boundary state of SEP and located in the tangential hyperplane, $\Tr\rho K=0$.
Here $\G$ is the abelian group of unitaries generated by the $(\one -P_{k,\ell})/2$.
$\langle\,\,\rangle_\G$ denotes symmetrizing by the ``twirl'' operation
concerning the group $\G$.
\end{itemize}
\end{thm}

Some states have more symmetry and the tangential witnesses can be found in some even smaller set,
showing the same symmetry as the state.
Sometimes these extra symmetries are given as subgroups of the inner
symmetries of $\W$ which we analyzed in Section \ref{symmetries}.
The simplest example concerns the {\it isotropic} witness, which is the
optimal entanglement witness for a Bell state.
The elementary calculation may again be performed for general $d$ as it is done for $d=3$.
Sometimes external groups, mapping part of $\W$ to other states, have to be used.
We give one example.

\begin{thm}
Concerning the subsection $\{\rho=\sum_kc_kP_{k,0}\}$
of density matrices, the search for witnesses can be reduced to
$\{K=\sum_k\kappa_kP_{k,0}+\sum_\ell \gamma_\ell\, Q_\ell\}$,
with $Q_\ell=\sum_k P_{k,\ell}\,/d$, and $\gamma_{-\ell}=\gamma_\ell$.
\end{thm}
\begin{proof}
As a first step we use the $\W$-symmetry of space reflection $\ell\leftrightarrow -\ell$.
This is an invariance of the chosen states.
Projecting $K=\sum_{k,\ell} \kappa_{k,\ell} P_{k,\ell}$ onto an invariant operator by
the ``twirl'' operation with this group $\G$ of two elements gives
\beq\label{kappa}
\langle K \rangle_\G =\sum_{k,\ell} \kappa_{k,\ell}' P_{k,\ell}\, ,
\qquad \kappa_{k,\ell}' =\frac12 (\kappa_{k,\ell}+\kappa_{k,-\ell}).
\eeq
In the second step we use
the group $\G U$ of local unitaries $\check{U}\otimes\tilde U$ diagonal in the preferred basis,
$$ \check{U}:\, |s\rangle\mapsto e^{i\alpha(s)}|s\rangle,\quad \tilde U:\, |t\rangle\mapsto e^{-i\alpha(t)}|t\rangle.$$
We use the expansion
\beq\label{pexp}
P_{k,\ell}=\frac1d \sum_{t,r}w^{k(t-r)}|t-\ell,t\rangle\langle r-\ell,r|,
\eeq
and form the second projection by twirl onto operators invariant under this group,
\beq\label{doublek}
\langle\langle K\rangle_\G \rangle_{\G U}=\sum_{k,\ell} \kappa_{k,\ell}' \frac1d \sum_{t,r}
w^{k(t-r)}\langle \, |t-\ell ,t\rangle\langle r-\ell,r| \, \rangle_{\G U}.
\eeq
The invariant part involves
\beqa\nonumber\langle \, |t-\ell ,t\rangle\langle r-\ell,r| \, \rangle_{\G U} =
\langle e^{i(\alpha(t-\ell)-\alpha (t)-\alpha (r-\ell)+\alpha(r))}\,\rangle_{\G U} |t-\ell ,t\rangle\langle r-\ell,r| \\
= \delta_{\ell,0}|t ,t\rangle\langle r,r| \quad +\quad
(1-\delta_{\ell,0})\delta_{t,r} |t-\ell ,t\rangle\langle t-\ell,t| \nonumber
\eeqa
Inserting this equation and also (\ref{kappa}) into (\ref{doublek})
gives for the first term  $\sum_k\kappa_{k,0}P_{k,0}$,
for the second term $(1-\delta_{\ell,0})\gamma_\ell Q_\ell$ with
$\gamma_\ell=\frac12 \sum_k (\kappa_{k,\ell}+\kappa_{k,-\ell})$.
\end{proof}

\section{Partial transposition}\label{partialtrans}

PT can be used, referring to the Peres criterion, to prove entanglement.
On the other hand it maps LMM$\cap$PPT onto itself, see Lemma \ref{lmmpt}.
There is a PT related subset $\hat{\W}$ of LMM with $\hat{\W}\cap$PPT$=\W\cap$PPT:
It is defined as the linear extension of ${\rm PT}(\W \cap {\rm PPT})$
to the borders of positivity.
The dimensions of these related subspaces are equal, ${\rm dim}(\W )={\rm dim}(\hat{\W})=d^2-1$.
Studies on the structure of $\W$ are automatically studies on
the structure of $\hat{\W}$.
The two pictures Fig.2 and Fig.3 presented in \cite{BHN06} for $d=3$
can be seen in that way.
The region of PPT-matrices (not necessarily positive) becomes the region of
states, i.e. positive matrices, and vice versa.
Their intersections are the PPT-states -- density matrices which are both positive and PPT --
in both points of view. The Peres criterion, SEP$\subset$PPT, implies that also
$\W\cap$SEP$=\hat\W \cap$SEP.
The cases of {\it bound entanglement}, \cite{HHH98},
may therefore also be seen in two ways. The regions of bound entanglement in $\W$, e.g. those
that we found for $d=3$, are in one to one correspondence to those in $\hat\W$.

PT of our simplex $\W$ has nice features, inferring simplification for calculations.
We use again the expansion (\ref{pexp})
and observe that PT maps $$|t-\ell,t\rangle\langle r-\ell,r|\mapsto|m-t,t\rangle\langle m-r,r|
\quad {\rm with} \quad m=t+r-\ell .$$
Splitting the global Hilbert space into subspaces according to the quantum number $m$ allows
for a splitting of partial transposed $\W$-states:
\beq\label{pt1}
{\rm PT}\, :\quad \rho=\sum_{k,\ell}c_{k,\ell}P_{k,\ell}\quad \mapsto\quad \bigoplus_m B_m\, ,
\eeq
with hermitian $d\times d$ matrices $B_m$,
\beq
\langle s|B_m|t\rangle=\frac1d\sum_k c_{k,s+t-m}w^{k(s-t)}=\langle t|B_m|s\rangle^\ast\, .
\eeq
\begin{thm}\label{ptequi}
Consider the matrices $B_m$ corresponding to some state in $\W$ according to (\ref{pt1}).
\begin{itemize}
\item For odd $d$ all the $B_m$ are unitarily equivalent.
\item For even $d$ there are two classes of mutually equivalent $B_m$,
one for even $m$, the other for odd $m$.
\item If $d$ is even, there is the relation of matrix elements for every $B_m$
\beq\label{pt2}
\langle s+d/2|B_m|t+d/2\rangle=\langle s|B_m|t\rangle .
\eeq
\end{itemize}
\end{thm}
\begin{proof}
For any $d$ observe
$$
\langle s|B_{m-2}|t\rangle=\frac1d\sum_k  c_{k,s+t-m+2}\,w^{k(s-t)}   = \langle s+1|B_m|t+1\rangle
$$
For $d$ odd one shows
$
\langle s|B_{m-1}|t\rangle=\langle s+(d+1)/2|B_m|t+(d+1)/2\rangle
$
by observing $s+t-m+1\equiv [s+(d+1)/2]+[t+(d+1)/2]-m$
in the second index of $c$.
For even $d$, the equivalence $s+t\equiv [s+d/2]+[t+d/2]$ implies (\ref{pt2}).
\end{proof}

The last point has the consequence that,
if $d$ is even, each $B_m$ has the form of a block matrix
\beq\label{block}
\left(  \begin{array}{cc} C & D \\ D & C \end{array} \right)\cong C\otimes \one (2)+D\otimes \sigma_x
\cong \frac{C+D}2\oplus\frac{C-D}2
\eeq
with hermitian blocks $C_m$ and $D_m$. $\one(\nu)$ is the $\nu \times\nu$ unit matrix.

Consider now the abelian algebras
\beq
\mathcal{A}(d):=\{\sum_{k,\ell}a_{k,\ell} P_{k,\ell},\,\, a_{k,\ell}\in\C\}\cong M_0(d^2,\C)
\eeq
emerging as a linear span of the special density matrices.
Using Theorem \ref{ptequi} and (\ref{block})
the results of mapping by PT are the following subalgebras of $M(d^2,\C)$:
\begin{itemize}
\item If $d$ is odd: \quad ${\rm PT}:\quad\mathcal{A}(d)\mapsto M(d,\C)\otimes \one (d)$,
\item If $d$ is even: \quad ${\rm PT}:\quad\mathcal{A}(d)\mapsto M(d/2,\C)\otimes M_0(4,\C)\otimes \one (d/2)$.
\end{itemize}
A consequence is a simplification for checking whether a state in $\W$ is PPT or not.
These states are mapped to linear functionals of PT$(\mathcal{A}(d))$, represented either, if $d$ is odd, by
hermitian matrices in $M(d,\C)$ or, if $d$ is even, by four hermitian
matrices in $M(d/2,\C)$.

A further consequence is an insight into the structure of the state space $\hat{\W}$:

\begin{thm}
The subset  $\hat{\W}$ of LMM is given by the intersection of ${\rm PT}(\mathcal{A}(d))$
with the set of density matrices.
\end{thm}

Only for $d=2$ it is again a simplex -- the reflected tetrahedron.
For odd $d$ it is the state space consisting of hermitian $d\times d$ density matrices
-- when the tensorial factor $\one$ is neglected.
For even $d\geq 4$ there are three-dimensional sections with the form of a tetrahedron
through every point in this $d^2-1$ dimensional convex body.
In other directions there exist sections of dimension $d^2/4-1$
with the structure of
the state space with $(d/2)\times(d/2)$ density matrices.
The space of states for $M(\nu ,\C)$ is the convex set of normalized positive $\nu\times \nu$ matrices.
Every maximal face is equivalent to the set of normalized $(\nu -1)\times (\nu -1)$ matrices.
So its faces have dimension $\nu (\nu -2)$  at most.
It follows that the surface of $\hat{\W}$ is curved in many directions, if $d\geq 3$.
Part of the border of $\hat{\W}$ is the border of PPT$\cap\W$.
This border is therefore also curved in many directions.

Both local unitary transformations and the global complex conjugation map PPT onto itself.
So the symmetries established in Section \ref{symmetries}
are symmetries of  PPT$\cap\W$ and of $\hat{\W}$ too.
Also the witnesses for $\W$ can be transported to witnesses of $\hat{\W}$ by PT.
This follows from the ``self-adjointness'' of PT as a transformation in the
Hilbert-Schmidt space:
$\Tr ({\rm PT}[\rho]K) =\Tr (\rho {\rm PT}[K])$.

Finally a look at special LMM states in $\hat{\W}$. No Bell states are in $\hat{\W}$, if $d\geq 3$.
There is a set of Werner states instead, and
$\hat\W$ includes as many Werner states with some given mixing as $\W$ contains Bell states:
$d^2$ of them are extremal with a density matrix which is a
$d(d-1)/2$ dimensional projector.
The {\it set} of maximally exposed LMM$\cap$SEP states are mapped by PT onto itself.
Their number in $\hat{\W}$ is thus again $N(d)$, the same as in $\W$,
see Theorem \ref{numberofextrems}, and Thm. \ref{seprhothm} in Section \ref{subtleties}.

\section{Optimality}\label{subtleties}

We follow the second trail which aims at proving not to have overlooked anything.

\begin{thm}\label{seprhothm}
There is a one to one correspondence between the maximally exposed states
in SEP$\cap\W$ and the lines or sublattices with d points, generated by abelian
subgroups of $\Ww$.
\end{thm}
\begin{proof}
One half of this theorem is proven in the Proposition \ref{seprho}.
On the other hand, SEP$\cap$LMM is inside the enclosure polytope.
In the large space of hermitian matrices the extremal points of this polytope
lie at the intersections of the witness hyperplanes $B_\alpha$ and the positivity borders $A_\beta$,
with $\alpha \in Q \subset \T$, $\beta\in\T \backslash Q$.
Restricting the space to the space of normalized matrices gives the condition $|Q|=d$.
This is the condition to get those vertices of the enclosure polytope which are inside of $\W$.
They all have exactly the same distance to $\omega$ as the maximally exposed separable states.
But not all of them are separable; only those, where $Q$ is a line or a sublattice.
For $d$ prime this has been stated in \cite{N06}.
For general dimension $d$ we define
\beq\label{epswit}
K=\one-(1+\varepsilon )\sum_{\alpha\in Q}P_{\alpha}=\one-(1+\varepsilon )d\cdot \rho_Q ,
\eeq
and claim that it is an entanglement witness if $\varepsilon$ is small
and if $Q$ is {\it not} a line or a sublattice.
To prove this claim we have to show that $\forall \,\,|\varphi,\psi\rangle$ the expectation value
of (\ref{epswit}) is not negative,
\beq
\langle\varphi,\psi|K|\varphi,\psi\rangle \geq 0.
\eeq
With $P_\alpha=W_\alpha |\Omega_{0,0}\rangle\langle \Omega_{0,0}|W_\alpha^\dag$
one gets
\beq\label{witexp}
\langle\varphi,\psi|K|\varphi,\psi\rangle=
\|\varphi\|^2\|\psi\|^2-(1+\varepsilon )\sum_\alpha|\langle\varphi,\psi|W_\alpha|\Omega_{0,0}\rangle|^2.
\eeq
We insert the definition (\ref{omega}) of $\Omega_{0,0}$;
$$\langle\varphi,\psi|W_\alpha|\Omega_{0,0}\rangle=
\sdird \sum_s\langle\varphi|\check{W}_\alpha|s \rangle\langle\psi|s \rangle=
\sdird\langle\varphi|\check{W}_\alpha|\tilde{\psi}\rangle,$$
with $|\tilde{\psi}\rangle:=\sum_s\langle s|\psi\rangle^\ast |s\rangle$.
So we have to check the non-negativity of
\beq\label{witk}
\|\varphi\|^2\|\tilde{\psi}\|^2-\frac{1+\varepsilon}d\sum_{\alpha\in Q}|\langle\varphi|\check{W}_\alpha|\tilde{\psi}\rangle|^2.
\eeq
Since $Q$ is not a sublattice, the Weyl operators which appear in the sum
do not all commute with each other. That means
$\check{W}_\alpha \check{W}_\beta =e^{i\gamma}\check{W}_\beta \check{W}_\alpha $
with $e^{i\gamma}\neq 1$ for some pairs of operators, and there is no common eigenvector.
For each pair of vectors there is at least one $\alpha$ such that
$|\langle\varphi|\check{W}_\alpha|\tilde{\psi}\rangle|<\|\varphi\|\|\tilde{\psi}\|$.
There is only a finite number of operators and a compact set of normalized vectors;
one has equicontinuity and uniform boundedness,
$$ \exists\varepsilon>0,\, {\rm s.t.}\, \forall \varphi,\tilde{\psi}:\,\,
\sum_{\alpha\in Q}|\langle\varphi|\check{W}_\alpha|\tilde{\psi}\rangle|^2<d\cdot(1-2\varepsilon)\|\varphi\|^2\|\tilde{\psi}\|^2.
$$
So the claim that $K$ defined in (\ref{epswit}) is a witness for some $\varepsilon$ is proven:
\beq
\langle\varphi,\psi|K|\varphi,\psi\rangle>
\|\varphi\|^2\|\psi\|^2 (\varepsilon-2\varepsilon^2).
\eeq
Since $\Tr K\rho_Q=-\varepsilon$, the state $\rho_Q$ is shown to be entangled.
\end{proof}

{\it Remark:} The procedure connecting the expectations (\ref{witexp}) with the
formula (\ref{witk}) is used also  in Section \ref{witnesses}, Theorem \ref{charwitness}.

In Theorem \ref{seprhothm} we have proved that the geometric symmetry of the kernel polytope is smaller
than that for the enclosure polytope. One implication is:

\begin{lem}
Every E-compatible point transformation $M$ must be
a linear invertible mapping $\T\rightarrow\T$.
\end{lem}

\begin{proof}
The set of kernel vertices has to be mapped onto itself.
This set corresponds to the set of lines and sublattices with exactly $d$ points
in the phase space $\T$.
Every pair of phase space points lies on one line at least,
many pairs on not more than one.
Since the mappings are one to one, each of these one-line-only pairs
has to be mapped onto an equivalent one-line-only pair.
There are enough of them, like $[(p,q),(p+k ,q+1)]$,
to imply the {\it linearity}: every line is mapped onto a line.
\end{proof}

We remark that invertibility of $M$ means
that the Matrix
$$M^{-1}=(\det M)^{-1}\left(  \begin{array}{cc} n & -m \\ -\ell & k \end{array} \right)$$
has to exist. This is only then the case if $\det M$ is coprime with $d$,
excluding e.g. $\det M=2$ for $d=4$, and $\det M=\pm 2$ or $3$, for $d=6$.

Next we show that the geometric symmetry of the kernel polytope is still
deceptive, if $d=5$ or $d\geq 7$.
\begin{thm}\label{esymplecticthm}
Consider a linear mapping $\T \rightarrow \T$ defined by applying a
$2\times 2$ matrix $M$ with elements $\in \Z_d$.
Its extension to a mapping $\W \rightarrow \W$ is E-compatible if and only if $\,\det M =\pm 1$.
\end{thm}

\begin{proof}
The first part is proven constructively in the proof of Proposition \ref{esymplectic}.
To prove the other direction we use duality of convex cones.
A linear mapping $\W\mapsto\W$ is E-compatible iff the dual transformation
maps SW to SW and TW onto TW.
The dual transformation, acting onto witnesses,
is given by the dual mapping of the set $\{\kappa_{k,\ell}\}$
considered as an element of $\ell^2 (\T ,\R )$:
$$K=\sum \kappa_{k,\ell}P_{k,\ell},\quad
\rho=\sum c_{k,\ell}P_{k,\ell}\quad\Rightarrow\quad
\Tr K\rho =\sum \kappa_{k,\ell}\,c_{k,\ell},\quad$$
The dual mapping of $\T$ is therefore $M^{-1}$, which is an element
of Sp(2,$\Z_d$) iff $M$ is such a matrix.

Consider now a line of tangential witnesses
\beq
K(\eps )=\lambda(\eps)\one+H+\eps P
\eeq
The parameter $\lambda$ is fixed through the conditions on $K$ stated in Theorem \ref{charwitness}.
They imply the existence of normed vectors $|\varphi ,\psi\rangle$ such that
\beq
\langle\varphi ,\psi|K|\varphi ,\psi\rangle = \min_{\chi,\eta}\langle\chi ,\eta|K|\chi ,\eta\rangle =0,
\eeq
and therefore
\beq
-\lambda (\eps)= \min_{\chi,\eta}\langle\chi ,\eta|(H+\eps P)|\chi ,\eta\rangle  .
\eeq
This situation is treated perturbatively.
Let $|\varphi(\eps) ,\psi(\eps)\rangle$ be a differentiable curve of vectors with
$|\varphi(0) ,\psi(0)\rangle=|\varphi ,\psi\rangle$, the minimizers at $\eps=0$, with normalized vectors
\quad$\varphi(\eps)=\varphi+\eps\delta\varphi+O(\eps^2)$,
\quad$\psi(\eps)=\psi+\eps\delta\psi+O(\eps^2)$. With
$$-\mu(\eps)=\langle\varphi(\eps) ,\psi(\eps)|(H+\eps P)|\varphi(\eps) ,\psi(\eps)\rangle$$
one gets
\beq\label{perturb}
-\frac d{d\eps}\mu(\eps)\mid_{\eps =0}\, =
\langle\varphi ,\psi|P|\varphi ,\psi\rangle
+[\langle\delta\varphi|\check{H}_\psi|\varphi\rangle + c.c.]
+[\langle\delta\psi|\tilde{H}_\varphi|\psi\rangle + c.c.],
\eeq
with the operators $\check{H}_\psi$ and $\tilde{H}_\varphi$ defined as quadratic forms in the local Hilbert spaces,
$$\langle\chi|\check{H}_\psi|\eta\rangle :=\langle\chi , \psi|H|\eta,\psi\rangle\, , \qquad
\langle\chi|\tilde{H}_\varphi|\eta\rangle :=\langle\varphi , \chi|H|\varphi,\eta\rangle \, .$$
We know from standard perturbation theory that the terms in square brackets in (\ref{perturb}) are
zero if the ground states of the local $\check{H}_\psi$ and $\tilde{H}_\varphi$ are not degenerate.

For $\eps=0$ we choose
$$H=\sum_k\gamma_k P_{k,0} \,\quad \gamma_k=-(w^k+w^{-k}).$$
Using the expansion (\ref{pexp}) and then $\frac 1d\sum_k\gamma_kw^{k(s-t)}=\delta_{s,t-1}+\delta_{s,t+1}$
we get
\beqa\label{hexpec}
\langle\varphi ,\psi|H|\varphi ,\psi\rangle &=& -\sum_{s,t}\varphi_s\psi_s\varphi_t^\ast\psi_t^\ast
(\delta_{s,t-1}+\delta_{s,t+1})\\
&=&-2\sum_s f(s)f^\ast(s+1),\label{hexef}
\eeqa
with $f(s):=\varphi_s\psi_s$.
The minimum of  (\ref{hexef}) is attained
if all $f(s)$ are real valued and positive.
Also the $\varphi_s$ and $\psi_s$ can be chosen as positive.
One step in minimizing (\ref{hexpec}) with the condition $\|\varphi\|=\|\psi\|=1$
can be considered as equivalent to the reverse task
of keeping the $f(s)$
fixed, and minimizing $\|\varphi\| \cdot\|\psi\|$.
This gives $\varphi_s=\psi_s$ and $\sum_sf(s)=\|\varphi\|^2=1$.
Using this as a side condition to minimize (\ref{hexef})
one gets the minimizers: for $d\geq 5$
they are $f(s)=\frac12 (\delta_{s,t}+\delta_{s,t+1})$
and $f(s)=\frac12 \delta_{s,t}+\frac14(\delta_{s,t-1}+\delta_{s,t+1})$ for any $t$.
Defining $\varphi=\sqrt{f}$ and $\psi=\sqrt{f}$
one sees the non-degeneracy of the ground states of the local operators $H_\psi$ and $\tilde{H}_\varphi$.
Using the ground state vectors $\varphi_s(\eps)$ and $\psi_s(\eps)$
gives $\mu (\eps)=\lambda(\eps)$.
Applying (\ref{perturb}) results therefore in
\beq\label{final}
-\frac d{d\eps}\lambda(\eps)\mid_{\eps =0}\, =
\langle\varphi ,\psi|P|\varphi ,\psi\rangle.
\eeq
Choose $\varphi_s=\psi_s=\frac1{\sqrt{2}}(\delta_{s,0}+\delta_{s,1})$,
consider $P=P_{0,1}$ and transformations by
$$M=\left(  \begin{array}{cc} 1 & 0 \\ 0 & n \end{array} \right).$$
$H(0)$ is invariant under this transformation, but the perturbative
$P_{0,1}$ changes to $P_{0,n}$.
The value of (\ref{final}) is changed, unless $n=\pm 1$,
proving the non-invariance of TW.
Using transformations by symplectic matrices, every matrix $M$ can be transformed to this diagonal form
without changing its determinant.
So $\det M=\pm 1$ is a necessary condition.
\end{proof}

\section{Summary and Outlook}\label{summary}

The article extends on previous work \cite{BHN06}
for qutrits, but here the results are stated in a more mathematical context
and are generalized to arbitrary Hilbert space dimensions $d$.
We consider the state space of two qudits and analyze a certain subset, the simplex
(generalized tetrahedron) $\W$ as the main object of our investigations.
It is obtained starting from a certain maximally entangled pure state, a Bell type
state. By applying on one side the Weyl operators
other orthogonal Bell type states are formed, and the set of mixtures, the complex hull,
is the simplex $\W$.
The Weyl operators are related to a discrete classical phase space,
representing in turn the algebraic relations of the
Weyl operators. This analogy enables us to describe the local transformations of
the quantum state space of interest and is very useful for several proofs
in this paper:
Transformations of $\W$ onto itself can be considered as
transformations of the discrete classical phase space. Thus the
symmetries and equivalences can be studied by this means.

The simplex $\W$
is embedded in a $d^2$--dimensional Euclidean space equipped with
a norm (the Hilbert-Schmidt norm) and an inner product.
We analyze in detail the properties how it is embedded in the whole state space of two qudits and
discuss symmetries and equivalences inside the simplex $\W$, its facets and witnesses.
This is obtained via the Weyl group which is a kind of
``quantization'' of classical phase space.

Then we investigate the question of the geometry of separability. We
start with the construction of two polytopes, an inner (kernel polytope)
and an outer (enclosure polytope) fence for separability. They
define entanglement witnesses but are in general not optimal.
The outer fence, the enclosure polytope, has the same geometric
symmetry as $\W$. Because
we are able to construct \textit{optimal} entanglement
witnesses explicitly we obtain in principle the border between separable and entangled
states, sometimes even in analytic form.

With our method we find also the set of
\textit{bound} entangled states of the parameter subspace under
investigation, by applying the partial
transposition on one subsystem, which detects entanglement via PPT.
The obtained PPT-witnesses are sometimes different from the
entanglement witnesses for the
density matrices under consideration.
We stated and explored also a kind of ``duality''
where the partial transposition maps
PPT$\cap\W$ to PPT$\cap \hat \W$,
where $\hat \W$ is another convex subset of LMM (the set of locally maximally mixed states),
and the cases of bound entanglement detected
in $\W$ are also cases for bound entanglement in $\hat \W$.

Summarizing, we could present a detailed geometric structure of the subset
of bipartite qudits under investigation. We think that this will
help to find a good characterization of the whole state space and to
investigate measures for entanglement for higher dimensional
systems.

In the outlook, we hope the paper advances our knowledge of these structures, of
the convex hull of higher-dimensional generalizations of the two-qubit Bell states.
These two-qubit states have many applications in quantum information theory, and so their
characterization for higher dimension is a desirable research goal.
The higher dimensional generalizations also may have applications in quantum information, and display
interesting geometrical features on their own.
Since we show how to construct optimal witnesses explicitly and
how to determine regions where there is bound
entanglement, i.e. entanglement which cannot be distilled by local
operation and classical communication (LOCC),
these methods might be useful in quantum cryptography.
Regarding possible applications of our results and methods in a wider context, we note that
the quasiclassical structure fits also exactly into the
conditions needed for teleportation and dense coding, e.g.
Refs.~\cite{BW92,W01}.

Furthermore, according to the opinion of one of the referees, due to the
Choi-Jamio{\l}kowski isomorphism between a class of bipartite states and maps \cite{ZB04} or channels \cite{HHH98b}
one may infer from the results of the present paper further
conclusions about the maps eg. (bistochastic) superpositive maps,
which correspond to separable states:
Using the  isomorphism
(for a recent exposition see e.g. \cite{B06} or \cite{L06})
we see that
the states of a $d \times d$ system satisfying one partial trace condition
$\Tr_A \rho=\omega_B$ represent stochastic maps
(completely positive, trace preserving linear maps)
while states satisfying simultaneously both conditions
$\Tr_A \rho=\omega_B$, $\Tr_B \rho=\omega_A$, represent bistochastic maps.
Hence the set LMM for bipartite systems is isomorphic to the set of bistochastic maps.
See also \cite{BDS01} for relations to remote state preparation.

Last but not least, exploration of entanglement properties in still more detail,
using the high symmetry of the chosen sets of states,
seems to be the nearest and next goal.


\begin{thebibliography}{BGM85}


\bibitem[BW92]{BW92} C.H. Bennett, S.J. Wiesner:
{\it Communication via one- and two-particle operators on Einstein-Podolsky-Rosen state},
Phys. Rev. Lett. {\bf 69}, 2881 -- 4 (1992)

\bibitem[B93]{B93} C.H. Bennett et al.:
{\it Teleporting an unknown quantum state via dual
classical and Einstein-Podolsky-Rosen channels}, Phys. Rev. Lett
{\bf 70}, 1895 (1993)

\bibitem[W01]{W01} R.F. Werner:
{\it All teleportation and dense coding schemes},
J. Phys. A {\bf 34}, 7081 -- 94 (2001)

\bibitem[VW00]{VW00} K. G. H. Vollbrecht, R. F. Werner:
{\it Entanglement measures under symmetry}, Phys. Rev. A
{\bf 64}, 062307 (2001); arXiv:quant-ph/0010095

\bibitem[HH96]{HH96} R. Horodecki, and M. Horodecki:
{\it Information-theoretic aspects of inseparability of mixed states};
Phys. Rev. A {\bf 54}, 1838 -- 43 (1996)

\bibitem[ZB04]{ZB04} K. Zyczkowski, I. Bengtsson:
{\it On Duality between Quantum Maps and Quantum States}, Open Syst. Inf. Dyn.
{\bf 11}, 3-42 (2004); arXiv:quant-ph/0401119

\bibitem[AU82]{AU82} P. M. Alberti, and A. Uhlmann:
{\it Stochasticity and Partial Order: Doubly Stochastic Maps and Unitary Mixing};
Mathematics and its applications/9.
VEB Deutscher Verlag der Wissenschaften, Berlin 1981, and
(M. Hazewinkel editor)
D.Reidel, Publ. Company, Dordrecht (1982)

\bibitem[BNT02]{BNT02} R.A. Bertlmann, H. Narnhofer and W. Thirring:
{\it Geometric picture of entanglement and Bell inequalities}, Phys. Rev. A
{\bf 66}, 032319 (2002); arXiv:quant-ph/0111116

\bibitem[BHN06]{BHN06} B. Baumgartner, B. Hiesmayr, H. Narnhofer:
{\it The state space for two qutrits has a phase space structure in its core}, UWThPh-2006-13,
Phys. Rev. A {\bf 74}, 032327; arXiv:quant-ph/0606083

\bibitem[P96]{P96} A. Peres:
{\it Separability Criterion for Density Matrices},
Phys. Rev. Lett. {\bf 77}, 1413 (1996);
arXiv:quant-ph/9604005

\bibitem[T00]{T00} B.M. Terhal:
{\it Bell inequalities and the separability criterion},
Phys. Lett. A {\bf 271}, 319 -- 26 (2000);
arXiv:quant-ph/9911057

\bibitem[Sch86]{Sch86}
Schroeder, Manfred R. :
{\it Number theory in science and communication: with applications in cryptography, physics,
digital information, computing and self-similarity},
Springer, Berlin, 1986 . (Springer series in information sciences ; 7 )

\bibitem[V64]{V64}
Valentine, Frederick A.:
{\it Convex sets},
McGraw-Hill, NY, 1964.
(McGraw-Hill series in higher mathematics)

\bibitem[A42]{A42}
Aitken, Alexander C. :
{\it Determinants and matrices},
Oliver and Boyd , Edingburgh, 1942
(University mathematical texts ; 1 )

\bibitem[W06]{W06}
Wikipedia, the free encyclopedia;
http://en.wikipedia.org/wiki/

\bibitem[GB02]{GB02} L. Gurvits, H. Barnum:
{\it Largest separable balls around the maximally mixed bipartite quantum state},
Phys. Rev. A {\bf 66}, 062311 (2002)

\bibitem[W31]{W31} H. Weyl:
{\it Gruppentheorie und Quantenmechanik}, zweite Auflage,
(S. Hirzel, Leipzig, 1931)

\bibitem[B04]{B04} I. Bengtsson:
{\it MUBs, polytopes and finite geometries};
arXiv:quant-ph/0406174

\bibitem[N06]{N06} H. Narnhofer:
{\it Entanglement reflected in Wigner Functions},
J. Phys. A {\bf 39}, 7051 -- 64 (2006);

\bibitem[G99]{G99} D. Gottesman:
{\it Fault-Tolerant Quantum Computation with Higher-Dimensional Systems},
Lect. Notes. Comp. Sci. {\bf 1509}, 302 (1999); arXiv:quant-ph/9802007

\bibitem[GKP01]{GKP01} D. Gottesman, A. Kitaev, J. Preskill:
{\it Encoding a qubit in an oscillator},
Phys. Rev. A {\bf 64}, 012310, (2001)

\bibitem[HHH98]{HHH98} M. Horodecki, P. Horodecki and R. Horodecki:
{\it Mixed-State Entanglement and Distillation:
Is there a ``Bound'' Entanglement in Nature?},
Phys. Rev Lett. {\bf 80}, 5239 -- 42 (1998)

\bibitem[HHH98b]{HHH98b} M. Horodecki, P. Horodecki and R. Horodecki:
{\it General teleportation channel, singlet fraction and quasi-distillation},
arXiv:quant-ph/9807091

\bibitem[B06]{B06} I. Bengtsson:
{\it Geometry of quantum states},
(Cambridge, 2006)

\bibitem[L06]{L06} M.S. Leifer:
{\it Conditional Density Operators and the Subjectivity of Quantum Operations},
arXiv:quant-ph/0611233

\bibitem[BDS01]{BDS01} C.H. Bennett, D.P. DiVincenzo, P.W. Shor, et al :
{\it Remote State Preparation},
Phys. Rev Lett. {\bf 87}, 077902 (2001)


\end{thebibliography}
\end{document}